\documentclass{iaus}
\usepackage{graphicx}

\title[The MiMeS Project]{The MiMeS Project: \\Overview and current status}   
\author[G.A. Wade et al.]{G.A. Wade$^1$, E. Alecian$^2$, D.A. Bohlender$^3$, J.-C. Bouret$^4$, D. Cohen$^5$, V. Duez$^6$, M. Gagn\'e$^7$, J. Grunhut$^1$, H.F. Henrichs$^8$, N. Hill$^9$, O. Kochukhov$^{10}$, S. Mathis$^{11}$, C. Neiner$^{12}$, M. Oksala$^{13}$, S. Owocki$^{13}$, V. Petit$^7$, M. Shultz$^1$, T. Rivinius$^{14}$, R. Townsend$^9$, J.S. Vink$^{15}$\\and the MiMeS Collaboration\thanks{www.physics.queensu.ca/$\sim$wade/mimes}}   
\affiliation{$^1$Kingston, Canada, $^2$LOAG, France, $^3$HIA, Canada, $^4$LAM, France, $^5$Swarthmore, USA, $^6$Argelander, Germany, $^7$West Chester, USA, $^8$Amsterdam, Netherlands, $^9$Madison, USA, $^{10}$Uppsala, Sweden,$^{11}$CEA, France, $^{12}$Paris, France, $^{13}$Delaware, USA, $^{14}$ESO, Chile, $^{15}$Armagh, UK}    

\pubyear{2010}
\volume{272}  
\pagerange{119--126}
\jname{Active OB stars}
\editors{A.C. Editor, B.D. Editor \& C.E. Editor, eds.}
\begin{document}

\maketitle

\begin{abstract}
The Magnetism in Massive Stars (MiMeS) Project is a consensus collaboration among many of the foremost international researchers
of the physics of hot, massive stars, with the basic aim of understanding the origin, evolution and impact of 
magnetic fields in these objects. At the time of writing, MiMeS Large Programs have acquired over 950 high-resolution polarised spectra of about 150 individual stars with spectral types from B5-O4,
discovering new magnetic fields in a dozen hot, massive stars. The quality of this spectral and magnetic mat\'eriel is very high, and the Collaboration is keen to connect with colleagues capable of exploiting the data in new or unforeseen ways. In this paper we review the structure of the MiMeS observing programs and report the status of observations, data modeling and development of related theory.
\end{abstract}

\firstsection 
\section{Introduction}

Massive stars are those stars with initial masses above about 8 times that of the sun, eventually ending their lives in catastrophic supernovae. These represent the most massive and luminous stellar component of the Universe, and are the crucibles in which the lion's share of the chemical elements are forged. These rapidly-evolving stars drive the chemistry, structure and evolution of galaxies, dominating the ecology of the Universe - not only as supernovae, but also during their entire lifetimes - with far-reaching consequences. 

The Magnetism in Massive Stars (MiMeS) Project represents a comprehensive, multidisciplinary strategy by an international team of top researchers to address the Òbig questionsÓ related to the complex and puzzling magnetism of massive stars. MiMeS has been awarded "Large Program" status by the Canada-France-Hawaii Telescope (CFHT) and the T\'elescope Bernard Lyot (TBL), resulting in a total of 780 hours of time allocated to the Project with the high-resolution spectropolarimeters ESPaDOnS and Narval from late 2008 through 2012. This commitment of the observatories, their staff, their resources and expertise is being used to acquire an immense database of sensitive measurements of the optical spectra and magnetic fields of massive stars, which will be combined with a wealth of new and archival complementary data (e.g. optical photometry, UV and X-ray spectroscopy), and applied to address the 4 main scientific objectives of the MiMeS Project: 

\begin{enumerate}
\item To identify and model the physical processes responsible for the generation of magnetic fields in massive stars; 

\item To observe and model the interaction between magnetic fields and hot stellar winds; 

\item To investigate the role of the magnetic field in modifying the bulk rotation and differential rotational profiles of massive stars;

\item To understand the overall impact of magnetic fields on massive star evolution, and the connection between magnetic fields of non-degenerate massive stars and those of neutron stars and magnetars, with consequential constraints on stellar 
evolution, supernova astrophysics and gamma-ray bursts.  
\end{enumerate}

\section{Structure of the Large Programs}

\begin{figure}
 \includegraphics[width=13.5cm]{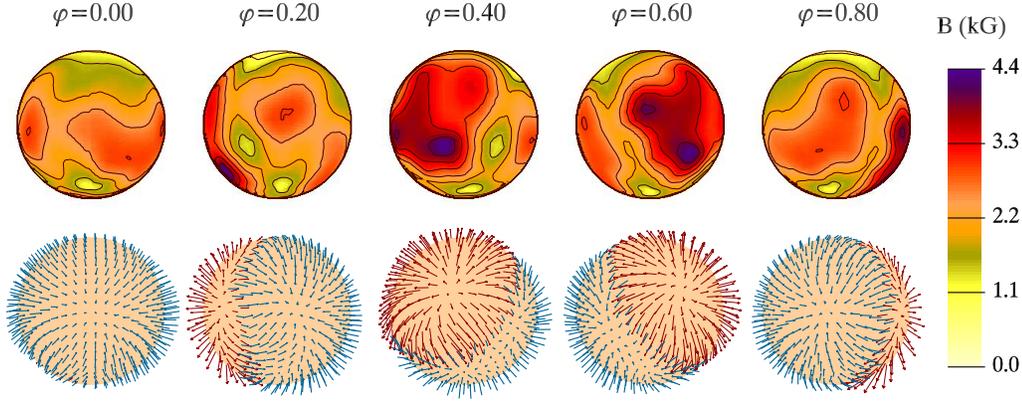}
  \caption{Magnetic Doppler Imaging (MDI) of the B9Vp star $\alpha^2$~CVn (HD 112413; Kochukhov \& Wade 2010), illustrating the reconstructed magnetic field orientation (lower images) and intensity (upper images) of this star at 5 rotation phases. The maps were obtained from a time-series of 21 Stokes $IVQU$ spectral sequences. Maps similar to these will be constructed for other stars in the MiMeS Targeted Component.}\label{fig:wave}
\end{figure}

To address these general problems, we have devised a two-component observing program that 
will allow us to obtain basic statistical information about the magnetic properties of the overall 
population of hot, massive stars (the Survey Component), while simultaneously providing detailed 
information about the magnetic fields and related physics of individual objects (the Targeted Component).

{\bf Targeted Component:} The MiMeS Targeted Component (TC) will provide data to map the magnetic fields and 
investigate the physical characteristics of a sample of known magnetic stars of great interest, 
at the highest level of sophistication possible. The roughly 25 TC targets have been selected to allow us to investigate a variety of 
physical phenomena, and to allow us to directly and quantitatively confront the predictions of stellar evolution theory, 
MHD magnetised wind simulations, magnetic braking models, etc. 

Each TC target is to be observed many times with ESPaDOnS and Narval, in order to obtain a high-precision and
high-resolution sampling of the rotationally-modulated circular (and sometimes linear) polarisation line profiles. Using state-of-the-art
tomographic reconstruction techniques such as Magnetic Doppler Imaging (Piskunov \& Kochukhov 2002), detailed maps
of the vector magnetic field on and above the surface of the star will be constructed (e.g. see Fig. 1). In combination with new and archival
complementary data, detailed analyses will be undertaken to model their evolutionary states, rotational evolution and wind structure and dynamics.

\begin{figure}
 \includegraphics[width=4.3cm]{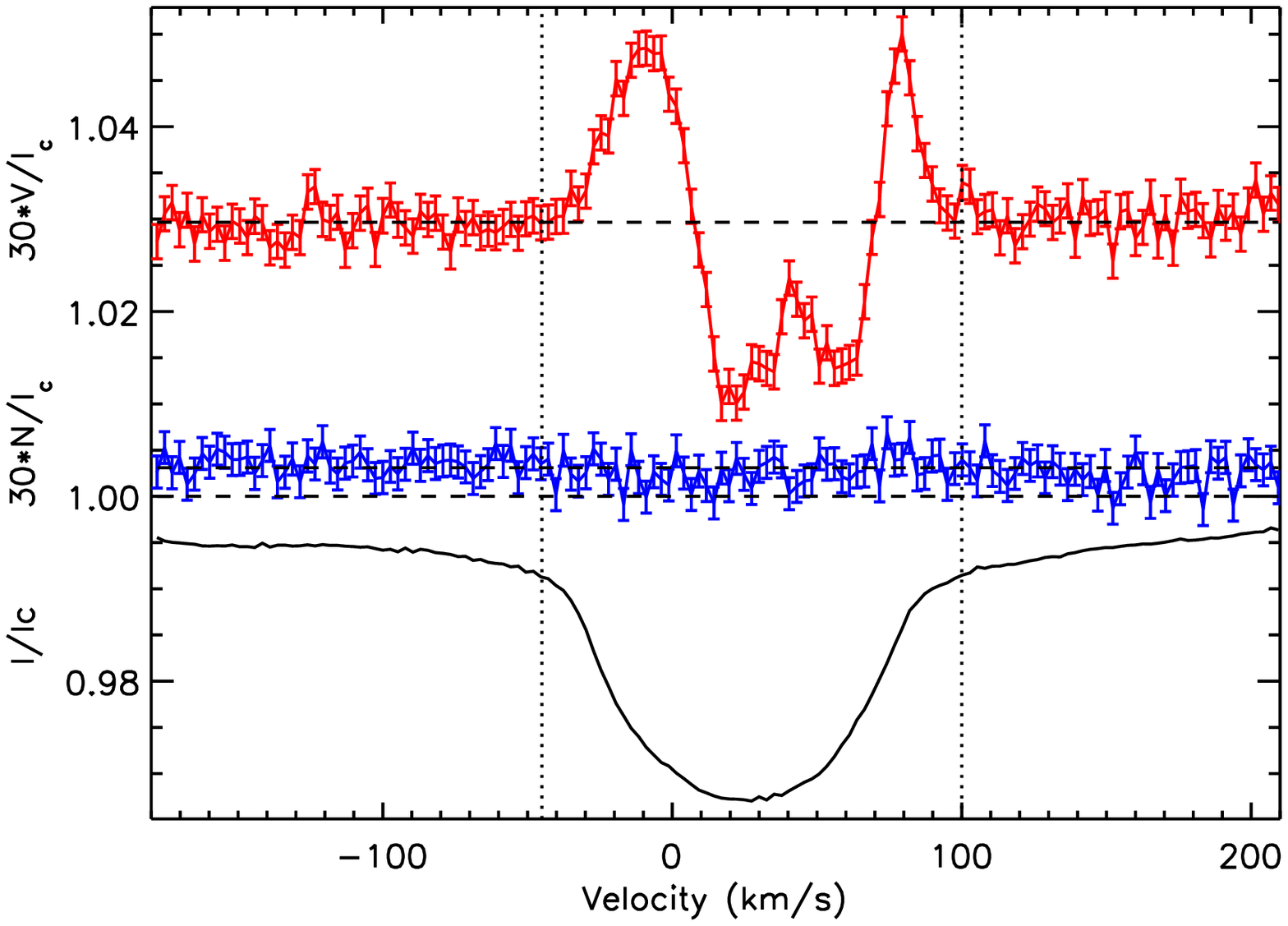}  \includegraphics[width=4.3cm]{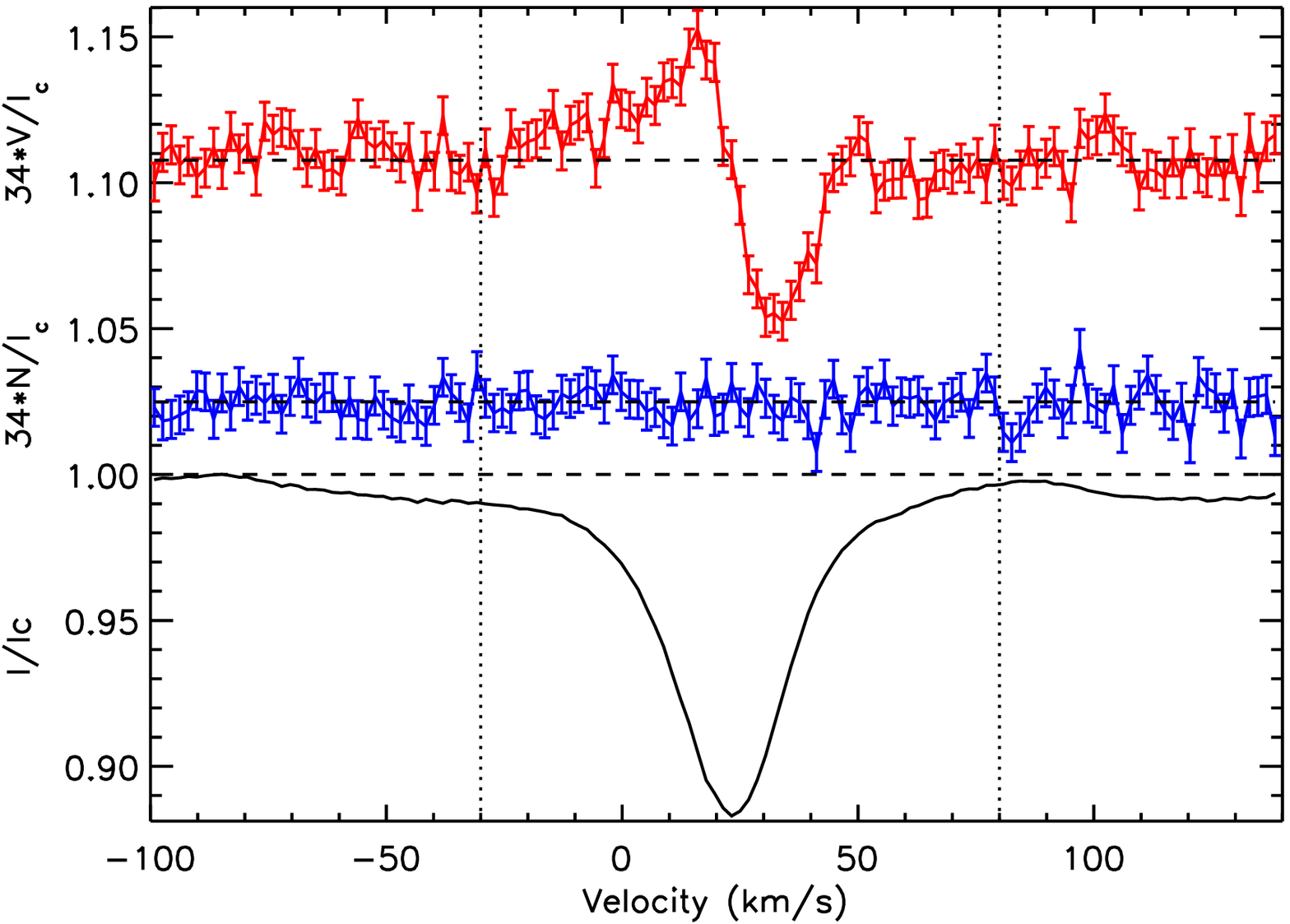} \includegraphics[width=4.3cm]{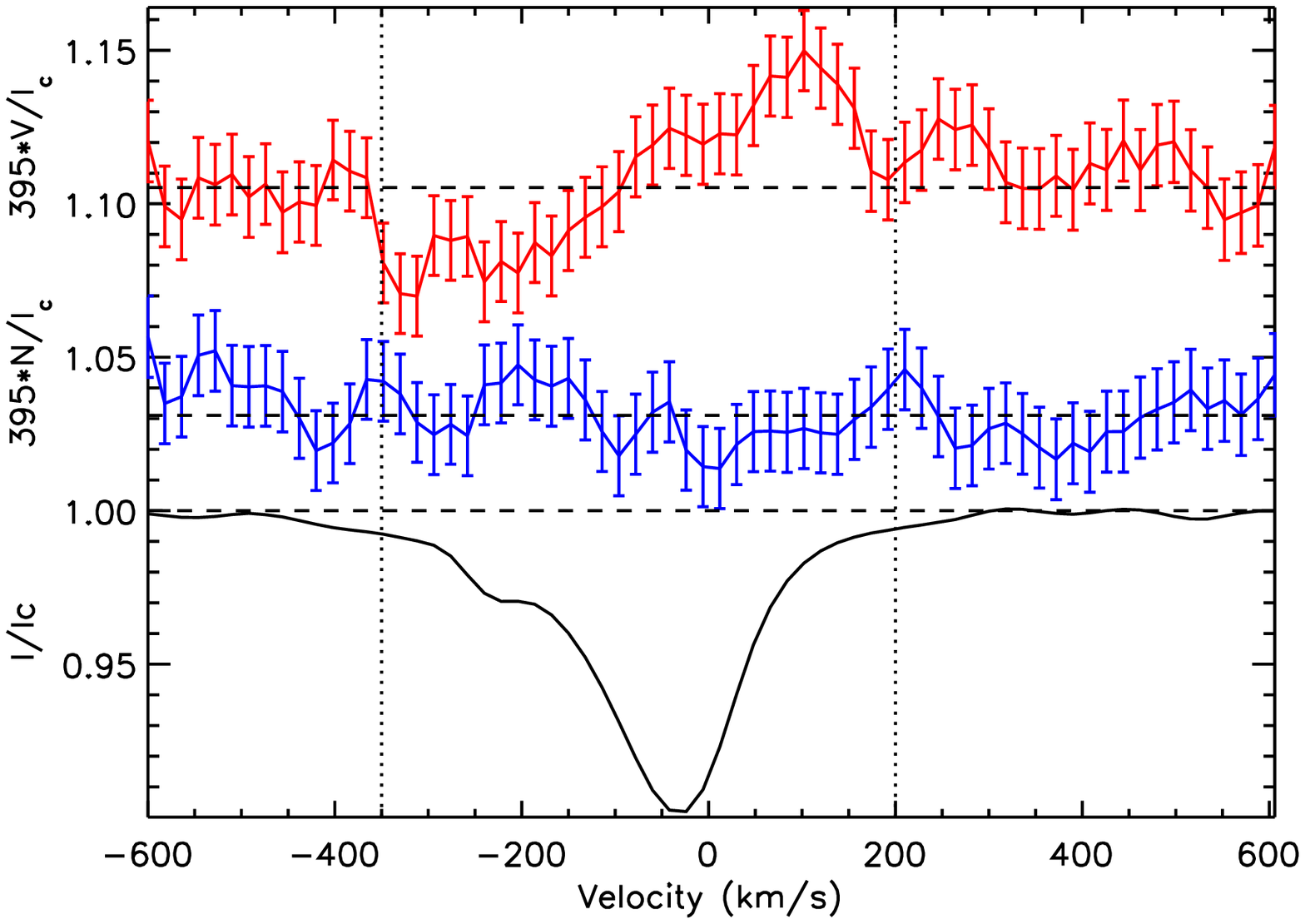}
  \caption{Least-Squares Deconvolved profiles of 3 hot stars in which magnetic fields have been discovered by the MiMeS Project: HD~61556 (B5V, left), HD 57682 (O9IV, middle) and HD 148937 (Of?p, right). The curves show the mean Stokes $I$ profiles (bottom curve), the mean Stokes $V$ profiles (top curve) and the $N$ diagnostic null profiles (middle curve). Each star exhibits a clear magnetic signature in Stokes $V$. To date, a dozen new magnetic stars have been discovered through the MiMeS Survey Component. }\label{fig:wave}
\end{figure}

{\bf Survey Component:} The MiMeS Survey Component (SC) provides critical missing information about field 
incidence and statistical field properties for a much larger sample of massive stars. It will also serve to 
provide a broader physical context for interpretation of the results of the Targeted Component.  
From an extensive list of potential OB stars compiled from published catalogues, we have 
generated an SC target sample of about 200 targets, covering the full range of spectral types from B4-O4, which 
are selected to sample broadly the parameter space of interest, while being well-suited to field detection. Our target list includes pre-main sequence Herbig Be stars,
field and cluster OB stars, Be stars, and Wolf-Rayet stars. 

Each SC target has been, or will be, observed once or twice during the Project, at very high precision in circular polarisation (e.g. see Figs. 2 \& 3). From the SC data we will
measure the bulk incidence of magnetic massive stars, estimate the variation of incidence versus 
mass, derive the statistical properties (intensity and geometry) of the magnetic fields of massive stars, 
estimate the dependence of incidence on age and environment, and derive the general statistical 
relationships between magnetic field characteristics and X-ray emission, wind properties, rotation, 
variability, binarity and surface chemistry diagnostics.  

Of the 780 hours allocated to the MiMeS LPs, about one-half is assigned to the TC and one-half to the SC. 

\section{Precision magnetometry of massive stars}

For all targets we exploit the longitudinal Zeeman effect in metal and helium lines to detect and measure magnetic 
fields in the line-forming region. Splitting of a spectral line due to a longitudinal magnetic field into oppositely polarised $\sigma$ components produces a variation of circular polarisation across the line (commonly referred to as a Ò(Stokes $V$) Zeeman signatureÓ or Òmagnetic signatureÓ; see Fig. 2.). The amplitude and morphology of the Zeeman signature encode information about the strength and structure of the global magnetic field. 
For some TC targets, we will also exploit the transverse Zeeman effect to constrain the detailed local structure of the field. Splitting of a spectral line by a transverse magnetic field into 
oppositely polarised $\pi$ and $\sigma$ components produces a variation of linear polarisation (characterized by 
the Stokes $Q$ and $U$ parameters) across the line (e.g. Kochukhov et al. 2004, review by Petit in these proceedings).

\subsection{Survey Component}

For the SC targets, the detection of magnetic field is diagnosed using the Stokes $V$ detection criterion described by Donati et al. (1997), and the "odds ratio" computed using the powerful Bayesian 
estimation technique of Petit et al. (2008). After reduction of the polarised spectra using the Libre-Esprit optimal extraction code (see Fig. 3), we employ the Least-Squares Deconvolution (LSD; Donati et al. 1997) multi-line 
analysis procedure to combine the Stokes $V$ Zeeman signatures from many spectral lines into a single high-S/N mean profile (see Fig. 2), enhancing our ability to detect subtle magnetic signatures. Least-Squares 
Deconvolution of a spectrum requires a Òline maskÓ to describe the positions, relative strengths and magnetic sensitivities of the lines predicted to occur in the stellar spectrum.  In our analysis we employ custom line masks that we tailor interactively to best reproduce the observed stellar spectrum, in order to maximise our sensitivity to weak magnetic fields.

The exposure duration required to detect a Zeeman signature of a given strength 
varies as a function of stellar apparent magnitude, spectral type and projected rotational velocity. This 
results in a large range of detection sensitivities for our targets. The SC exposure times are based on 
an empirical exposure time relation derived from real ESPaDOnS observations of OB stars, and takes into account detection sensitivity gains resulting from LSD and velocity 
binning, and sensitivity losses from line broadening due to rapid rotation.  
Exposure times for our SC targets correspond to the time required to 
definitely detect (with a false alarm probability below $10^{-5}$) the Stokes $V$ Zeeman signature produced 
by a surface dipole magnetic field with a specified polar intensity. Although our calculated exposure times correspond to definite detections of a dipole magnetic field, 
our observations are also sensitive to substantially more complex field topologies.

Examples of SC targets in which magnetic fields have been discovered by MiMeS are shown in Fig. 2. 

\begin{figure}
 \includegraphics[width=13.5cm]{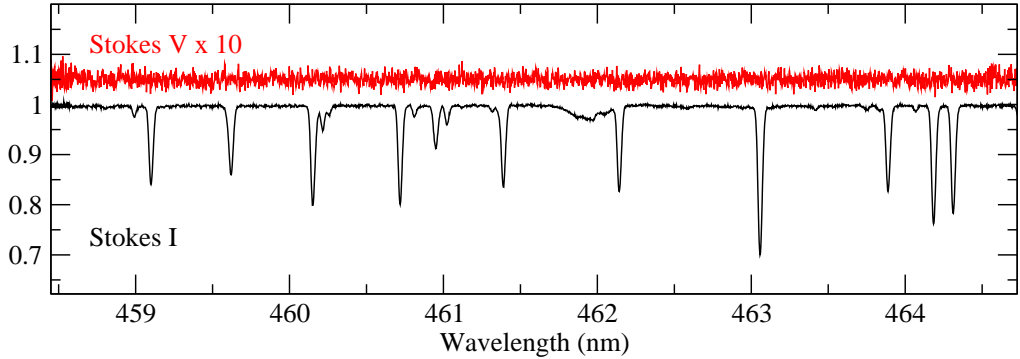}
  \caption{Detail of the MiMeS SC spectrum of the sharp-lined $\beta$~Cep star $\delta$~Ceti (= HD 16582). The peak S/N per 1.8 km/s pixel is 1000 (for an exposure time of 280 s), typical for an SC observation. LSD analysis yields no evidence of a magnetic field, with a 1$\sigma$ longitudinal field error bar of just 10~G.}\label{fig:wave}
\end{figure}

\subsection{Targeted Component}

Zeeman signatures are detected repeatedly in all spectra of TC targets. The spectropolarimetric timeseries are
interpreted using several magnetic field modeling codes at our disposal. For those stars for which 
Stokes $V$ LSD profiles will be the primary model basis, modeling codes such as those of Donati et al. 
(2006) or Alecian et al. (2008b) will be employed.  For those stars for which the signal-to-noise ratio in 
individual spectral lines is sufficient to model the polarisation spectrum directly, we will employ the Invers10 Magnetic Doppler Imaging code to 
simultaneously model the magnetic field, surface abundance structures and pulsation velocity field 
(Piskunov \& Kochukhov 2002, Kochukhov et al. 2004). The resultant magnetic field models will be 
compared directly with the predictions of fossil and dynamo models (e.g. Braithwaite 2006, 2007, Duez \& Mathis 2010, Arlt 2008).  

Diagnostics of the wind and magnetosphere (e.g. optical 
emission lines and their linear polarisation, UV line profiles, X-ray photometry and spectroscopy, radio 
flux variations, etc.) are being modeled using both the semi-analytic Rigidly-Rotating Magnetosphere 
approach, the Rigid-Field Hydrodynamics (Townsend et al. 2007) approach and full MHD simulations using the ZEUS 
code (e.g. Stone \& Norman 1992; ud Doula et al. 2008; Townsend, Hill, Rivinius, Grunhut in these proceedings; see also Fig. 4). 

\begin{figure}
 \includegraphics[width=13.5cm]{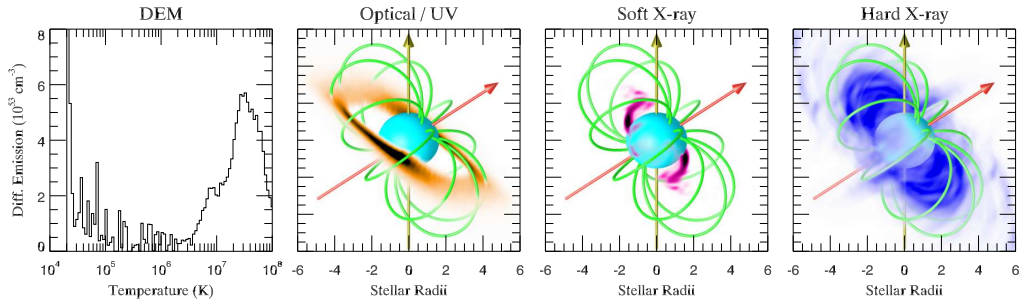}
  \caption{Example of the spectral and spatial emission properties of a rotating massive star magnetosphere modeled using Rigid Field Hydrodynamics (Townsend et al. 2007). The stellar rotation 
axis (vertical arrow) is oblique to the magnetic axis (inclined arrow), leading to complex plasma flow produced by radiative acceleration, Lorentz forces and centripetal acceleration. The consequent heated plasma distribution in the stellar magnetosphere (illustrated in colour/grey scale) shows broadband emission, and is highly structured both spatially and spectrally. Magnetically confined winds such as this are responsible for the X-ray emission and variability properties of some OB stars, and models such as this are being constructed for the MiMeS Targeted Component.
}\label{fig:wave}
\end{figure}

\section{Project status}

At the time of writing, MiMeS Large Programs have acquired over 950 spectra of over 150 individual stars. About 45\% of the spectra correspond to the SC, while 65\% correspond to TC targets. Following their acquisition in Queued Service Observing mode at the CFHT, ESPaDOnS polarised spectra are immediately reduced by CFHT staff using the Upena pipeline feeding the Libre-Esprit reduction package and downloaded to the dedicated MiMeS Data Archive at the Canadian Astronomy Data Centre (CADC) in Victoria, Canada. Narval data are similarly reduced by TBL staff then downloaded to the MiMeS Collaboration's Wiki site at Observatoire de Paris, France. All reduced spectra are carefully normalized to the continuum using custom software tailored to hot stellar spectra. Each reduced spectrum is then subject to an immediate quick-look analysis to verify nominal resolving power, polarimetric performance and S/N. The quality of the $\sim 420$ SC spectra acquired to date is very high: the median peak SNR of the Stokes $I$ spectra is 1150 (per 1.8 km/s spectral pixel) and the median longitudinal field error bar ($1\sigma$) derived from the Stokes $V$ spectra is just 37 G. A typical SC spectrum is illustrated in Fig. 3.

Preliminary LSD profiles are extracted using our database of generic hot star line masks to perform an initial magnetic field diagnosis and further quality assurance. Ultimately, each spectrum will be processed by the MiMeS Massive Stars Pipeline (MSP; currently in production) to determine a variety of critical physical data for each observed target, in addition to the precision magnetic field diagnosis: effective temperature, surface gravity, mass, radius, age, variability characteristics, projected rotational velocity, radial velocity and binarity, and mass loss rate. These meta-data, in addition to the reduced high-quality spectra, will be uploaded for publication to the MiMeS Legacy Database\footnote{The MiMeS Project is undertaken within the context of the broader MagIcS (Magnetic Investigations of various Classes of Stars) collaboration, www.ast.obs-mip.fr/users/donati/magics).}.

A large variety of MiMeS results are presented in these proceedings, both observational (Fourtune-Ravard et al., Grunhut et al., Henrichs et al., Oksala et al., Petit et al., Rivinius et al., Shultz et al. and Wade et al.) and theoretical (Duez et al., Mathis et al., Townsend et al., Hill et al.).

\begin{acknowledgments}
The MiMeS CFHT Large Program (2008B-2012B) is supported by both Canadian and French Agencies, and was one of 4 such programs selected in early 2008 as a result of an extensive international expert peer review of many competing proposals. The MiMeS TBL Large Program (2010B-2012B) was allocated in the context of the competitive French Time Allocation process.

Based on observations obtained at the Canada-France-Hawaii Telescope (CFHT) which is operated by the National Research Council of Canada, the Institut National des Sciences de l'Univers of the Centre National de la Recherche Scientifique of France, and the University of Hawaii. Also based on observations obtained at the Bernard Lyot Telescope (TBL, Pic du Midi, France) of the Midi-Pyr\'en\'ees Observatory, which is operated by the Institut National des Sciences de l'Univers of the Centre National de la Recherche Scientifique of France.

The MiMeS Data Access Pages are powered by software developed by the CADC, and contains data and meta-data provided by the CFH Telescope.
\end{acknowledgments}

\end{document}